\documentclass[aps,prl,twocolumn,superscriptaddress,floatfix]{revtex4-1}

\usepackage{graphicx}
\usepackage{dcolumn}
\usepackage{bm}
\usepackage{amssymb}
\usepackage{microtype}
\usepackage{xfrac}
\usepackage{array}
\newcommand{\latin}[1]{{\it #1}}

\newcommand{\ket}[1]{\ensuremath{|#1\rangle}}
\newcommand{\bra}[1]{\ensuremath{\langle#1|}}
\newcommand{\angstrom}{\text{\normalfont\AA}}

\begin{document}

\title{Quantum Spin Ice Dynamics in the Dipole-Octupole Pyrochlore Magnet Ce$_2$Zr$_2$O$_7$}

\author{~J.~Gaudet}
\affiliation{Department of Physics and Astronomy, McMaster University, Hamilton, ON, L8S 4M1, Canada}

\author{~E.~M.~Smith}
\affiliation{Department of Physics and Astronomy, McMaster University, Hamilton, ON, L8S 4M1, Canada}

\author{~J.~Dudemaine}
\affiliation{D\'epartement de Physique, Universit\'e de Montr\'eal, 2900 Boul.\'Edouard-Montpetit, Montr\'eal, QC, H3T 1J4, Canada}

\author{~J.~Beare}
\affiliation{Department of Physics and Astronomy, McMaster University, Hamilton, ON, L8S 4M1, Canada}

\author{~C.~R.~C.~Buhariwalla}
\affiliation{Department of Physics and Astronomy, McMaster University, Hamilton, ON, L8S 4M1, Canada}

\author{~N.~P.~Butch}
\affiliation{Center for Neutron Research, National Institute of Standards and Technology, MS 6100 Gaithersburg, Maryland 20899, USA}

\author{~M.~B.~Stone}
\affiliation{Neutron Scattering Division, Oak Ridge National Laboratory, Oak Ridge, Tennessee 37831, USA}

\author{~A.~I.~Kolesnikov}
\affiliation{Neutron Scattering Division, Oak Ridge National Laboratory, Oak Ridge, Tennessee 37831, USA}

\author{~Guangyong~Xu}
\affiliation{Center for Neutron Research, National Institute of Standards and Technology, MS 6100 Gaithersburg, Maryland 20899, USA}

\author{~D.~R.~Yahne}
\affiliation{Department of Physics, Colorado State University, 200 W. Lake St., Fort Collins, CO 80523-1875, USA}

\author{~K.~A.~Ross}
\affiliation{Department of Physics, Colorado State University, 200 W. Lake St., Fort Collins, CO 80523-1875, USA}
\affiliation{Canadian Institute for Advanced Research, 180 Dundas St. W., Toronto, ON, M5G 1Z7, Canada}

\author{~C.~A.~Marjerrison}
\affiliation{Brockhouse Institute for Materials Research, McMaster University, Hamilton, ON L8S 4M1 Canada}

\author{~J. D.~Garrett}
\affiliation{Brockhouse Institute for Materials Research, McMaster University, Hamilton, ON L8S 4M1 Canada}

\author{~G.~M.~Luke}
\affiliation{Department of Physics and Astronomy, McMaster University, Hamilton, ON, L8S 4M1, Canada}
\affiliation{Canadian Institute for Advanced Research, 180 Dundas St. W., Toronto, ON, M5G 1Z7, Canada}

\author{~A.~D.~Bianchi}
\affiliation{D\'epartement de Physique, Universit\'e de Montr\'eal, 2900 Boul.\'Edouard-Montpetit, Montr\'eal, QC, H3T 1J4, Canada}
\affiliation{Regroupement Qu\'eb\'ecois sur les Mat\'eriaux de Pointe (RQMP), Quebec, QC H3T 3J7 Canada }

\author{~B.~D.~Gaulin}
\affiliation{Department of Physics and Astronomy, McMaster University, Hamilton, ON, L8S 4M1, Canada}
\affiliation{Canadian Institute for Advanced Research, 180 Dundas St. W., Toronto, ON, M5G 1Z7, Canada}
\affiliation{Brockhouse Institute for Materials Research, McMaster University, Hamilton, ON L8S 4M1 Canada}

\date{\today}

\begin{abstract} 

Neutron scattering measurements on the pyrochlore magnet Ce$_2$Zr$_2$O$_7$ reveal an unusual crystal field splitting of its lowest $J$~=~5/2 multiplet, such that its ground state doublet is composed of m$_J$~=~$\pm$~3/2, giving these doublets a dipole - octupole (DO) character with local Ising anisotropy. Its magnetic susceptibility shows weak antiferromagnetic correlations with $\theta_{CW}$~=~-~0.4(2)~K, leading to a naive expectation of an All-In, All-Out ordered state at low temperatures. Instead our low energy inelastic neutron scattering measurements show a dynamic quantum spin ice state, with suppressed scattering near $\vert${\bf Q}$\vert$~=~0, and no long range order at low temperatures. This is consistent with recent theory predicting symmetry enriched U(1) quantum spin liquids for such DO doublets decorating the pyrochlore lattice. Finally, we show that disorder, especially oxidation of powder samples, is important in Ce$_2$Zr$_2$O$_7$ and could play an important role in the low temperature behaviour of this material.

\end{abstract}


\maketitle


The rare-earth pyrochlore oxides R$_2$B$_2$O$_7$, where R$^{3+}$ and B$^{4+}$ consist generally of rare earth and transition-metal ions respectively, display a wealth of both exotic and conventional magnetic ground states. Their R$^{3+}$ ions decorate a network of corner-sharing tetrahedra, one of the archetypes for geometrical frustration in three dimensions.  Due to strong crystal electric field (CEF) effects, the nature of the magnetic interactions in such materials are strongly influenced by their single-ion physics~\cite{Greedan,RevHallas,Raureview}. A naive theoretical description of the magnetic interactions in rare-earth pyrochlores is generally performed by introducing an \latin{ad hoc} effective single-ion term in addition to Heisenberg exchange interactions. For example, Heisenberg antiferromagnetism with an effective Ising anisotropy leads to non-frustrated All-In, All-Out (AIAO) magnetic order, as seen in several heavy rare earth iridate pyrochlores~\cite{sagayama2013,disseler2014} and illustrated in the insert to Fig.1(a). Heisenberg ferromagnetism and an effective Ising anisotropy give rise to a classical spin ice ground state~\cite{bramwell2001spin}, as seen in (Ho,Dy)$_2$Ti$_2$O$_7$~\cite{harris1997geometrical,morris2009} and illustrated as the 2I2O local structure in the inset to Fig.1(a). However, to capture all the physics that can arise at low temperatures, the magnetic interactions should be projected into pseudo-spin operators acting solely on the low energy CEF states~\cite{curnoe2008structural,onoda2011effective,Ross2011,Savary2012,Rau2015HTO,Raureview}. This procedure has been applied for example in the Yb$^{3+}$~\cite{Ross2011,robert2015spin,thompson2017quasiparticle} and Er$^{3+}$~\cite{Savary2012,petit2014order,Hallas2017EPO,Petit2017}  XY pyrochlores where CEF effects give rise to effective $S$~=~1/2 quantum degrees of freedom that interact via anisotropic exchange interactions.

\begin{figure*}[tbp]
\linespread{1}
\par
\includegraphics[width=7.1in,height=1.85in]{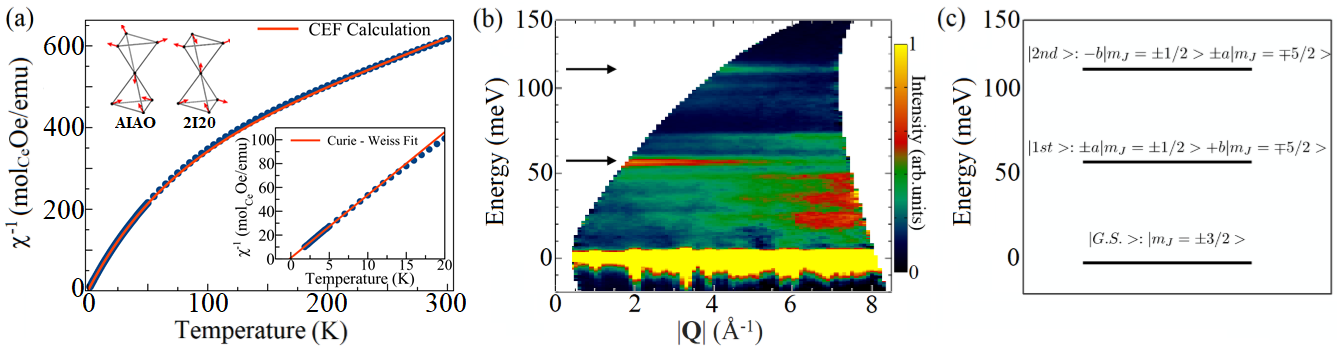}
\par
\caption{(a) The inverse magnetic susceptibility of a powder sample of Ce$_2$Zr$_2$O$_7$. The red curve is the Van-Vleck susceptibility calculated with the CEF Hamiltonian of Ce$_2$Zr$_2$O$_7$. The top left inset displays the AIAO and 2I2O magnetic ground state spin configurations on a pair of tetrahedra. The bottom right inset shows the low temperature magnetic susceptibility that yields $\theta_{CW}$~=~-0.4(2)~K and a paramagnetic moment of 1.3(1)~$\mu_B$, and shows no signature of magnetic order or spin freezing down to 0.5~K. (b) Inelastic neutron scattering spectra of Ce$_2$Zr$_2$O$_7$ at $T$~=~5~K with incident neutron energy $E_i$~=~150~meV. Two strong excitations can be identified as magnetic in origin at E~$\sim$~56 and $\sim$~112~meV, as their intensity decreases as a function of $\vert${\bf Q}$\vert$, consistent with the Ce$^{3+}$ magnetic form factor. This contrasts with phonon excitations whose intensity increases with $\vert${\bf Q}$\vert$, often as $\vert${\bf Q}$\vert^2$. (c) The energy eigenvalues corresponding to the CEF states belonging to the spin-orbit ground state manifold of Ce$_2$Zr$_2$O$_7$. The composition of the CEF eigenfunctions are also presented in (c), revealing the DO nature of the ground state doublet - that is, it corresponds to pure m$_J$~=~$\pm$~3/2 states. }
\label{CEF}
\end{figure*}

More recently, it has been realized that the precise composition of the ground state crystal field doublets in rare-earth pyrochlores is crucial in determining the form of the microscopic Hamiltonian, and in itself, diversifies the possibility of quantum magnetic states~\cite{Raureview,huang2014quantum}. This has been appreciated for some time in the case of non-Kramers doublets, based on magnetic ions with an even number of electrons such as the 4f$^2$ configuration in Pr$^{3+}$. Only the local $z$-component of the spin operators transforms as a dipole, with the transverse components transforming as quadrupoles~\cite{lee2012generic,chen2017,martin2017disorder}. This restricts the form of the effective spin Hamiltonian and can stabilize  quadrupolar phases that are not present in the phase diagram for dipolar doublets~\cite{onoda2011quantum,takatsu2016quadrupole}. For Kramers ions with an odd number of electrons, such as 4f$^1$ in Ce$^{3+}$, 4f$^3$ in Nd$^{3+}$ and 4f$^5$ in  Sm$^{3+}$, a crystal field ground state doublet with DO character can be realized where the local $z$ and $x$ components transform as a dipole, but the local $y$ component transforms as an octupole~\cite{huang2014quantum,li2017,lhotel2015,mauws2018}. After a rotation of the pseudo-spins about the $y$ axis, the DO exchange Hamiltonian on the pyrochlore lattice can be reduced to an XYZ model with three independent exchange parameters ($J_{\tilde{x}}$,$J_{\tilde{y}}$,$J_{\tilde{z}}$)~\cite{huang2014quantum,li2017}. This Hamiltonian allows for multiple phases to emerge such as AIAO order in the limit of isotropic antiferromagnetic interactions ($J_{\tilde{x}}\approx J_{\tilde{y}} \approx J_{\tilde{z}}$). The DO nature of the doublet allows for octupolar ordered phases~\cite{huang2014quantum,li2017} and also  for moment fragmentation within the ground state, as observed in Nd$_2$Zr$_2$O$_7$, where static AIAO order co-exists with dynamic spin ice fluctuations~\cite{benton2016quantum,petit2016}. In the limit of dominant antiferromagnetic interactions and strong easy-axis exchange anisotropy, a dipolar quantum spin ice is stabilized so long as the easy-axis is along one of the dipolar components of the DO doublet ($J_{\tilde{x}} >> J_{\tilde{z}}$,$J_{\tilde{y}}$ or $J_{\tilde{z}} >> J_{\tilde{x}}$,$J_{\tilde{y}}$). An octupolar quantum spin ice is favoured if the easy-axis is along the octupole component ($J_{\tilde{y}} > >J_{\tilde{x}}$,$J_{\tilde{z}}$)~\cite{huang2014quantum,li2017}. 

A promising family of candidate materials for dipolar or octupolar quantum spin ice physics originating from DO doublets are the cerium pyrochlores Ce$_2$B$_2$O$_7$. The Ce$^{3+}$ ions in the pyrochlore Ce$_2$Sn$_2$O$_7$ are believed to have a DO CEF ground state and to interact via dominant antiferromagnetic interactions, but do not magnetically order down to $T$~=~20~mK~\cite{sibille2015candidate,li2017}. The low energy spin dynamics of the cerium pyrochlores remains unexplored and their characterization is key in determining the nature of their possible spin liquid states. In this letter, we gain further insight into the novel magnetism of cerium pyrochlores by reporting new inelastic neutron scattering experiments on powder and single crystal samples of Ce$_2$Zr$_2$O$_7$. Using high energy inelastic neutron scattering, we first confirmed the DO nature of the Ce$^{3+}$ single ion ground state wave functions in Ce$_2$Zr$_2$O$_7$. We also present low energy inelastic neutron scattering measurements performed on a single crystal of Ce$_2$Zr$_2$O$_7$ and observe diffuse, inelastic magnetic scattering that emerges at low temperatures. The {\bf Q} dependence of this diffuse scattering is consistent with a symmetry-enriched U(1) quantum spin ice state at low but finite temperatures. Furthermore, we show the quantum spin-ice correlations remain dynamic down to at least 60~mK with no sign of static magnetic order. Our results demonstrate {\bf Q} signatures of a dynamic quantum spin ice ground state in Ce$_2$Zr$_2$O$_7$, with associated emergent quantum electrodynamics and elementary excitations based on magnetic and electric monopoles as well as emergent photons~\cite{hermele2004pyrochlore,savary2012coulombic,savary2013spin,bentonphoton}.

Single crystal and powder samples of Ce$_2$Zr$_2$O$_7$ have been grown using floating zone techniques and solid state synthesis. Stabilizing the Ce$^{3+}$ oxidation state in Ce$_2$Zr$_2$O$_7$ is not simple, and requires growth and annealing in strong reducing conditions to minimize Ce$^{4+}$~\cite{otobe2005}. As discussed in the Supplemental Material (SM), this is a serious issue, especially in powder samples, where oxidization is observed to occur in powders exposed to air on a time scale on the order of minutes, complicating the exact characterization of the material's stoichiometry. The oxidization process can be tracked through high resolution x-ray diffraction measurements of the lattice parameter, and it is much slower for single crystal samples. There we can make an estimate of the stoichiometry of the single crystal used in our experiments as Ce$_2$Zr$_2$O$_{7+\delta}$ with $\delta$~$\sim$~0.1.

We first present high energy inelastic neutron scattering measurements, which probe the degeneracy breaking of the Ce$^{3+}$ spin-orbit manifold that occurs due to CEF effects. To do so, we used the SEQUOIA high resolution inelastic chopper spectrometer~\cite{granroth2010sequoia} at the Spallation Neutron Source of Oak Ridge National Laboratory and employed neutrons with incident energies (E$_i$) of 150 and 500~meV. The E$_i$~=~150~meV instrument setting was chosen to resolve the crystal electrical field (CEF) states that belong to the spin-orbit ground state manifold ($J$~=~5/2). The CEF interaction lifts the Ce$^{3+}$ spin-orbit ground state degeneracy into three different eigenstates that are each doubly degenerate. We also estimated a CEF Hamiltonian for Ce$_2$Zr$_2$O$_7$ using a scaling procedure based on the Er$^{3+}$ pyrochlore CEF scheme~\cite{gaudet2018effect}. This predicts two CEF excited states near 80 and 100~meV with similar inelastic neutron scattering intensity at $T$~=~5~K. As seen in Fig.1(b), this scenario is in qualitatively good agreement with our 150~meV inelastic neutron experimental spectra where two strong magnetic excitations are observed at $\sim$~56 and $\sim$~112~meV. The relative scattered intensity of these CEF transitions can be obtained once minor corrections for a phonon background are made. This procedure is further described in the SM and gives $I_{56meV}$/$I_{112meV}$ = 1.2(1), in good agreement with our expectations based on this scaling argument.   

\begin{figure}[tbp]
\linespread{1}
\par
\includegraphics[width=3.4in]{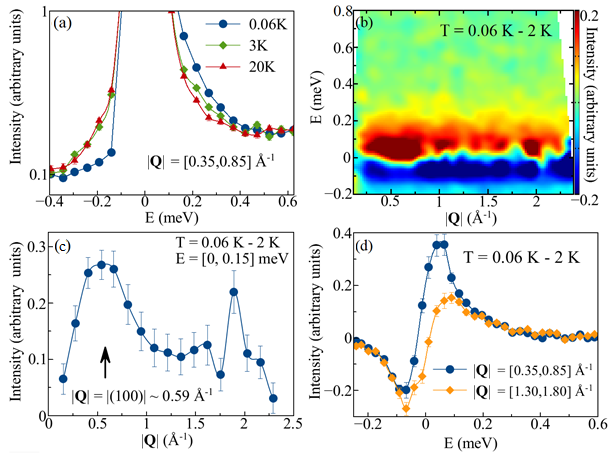}
\par
\caption{(a) The onset of dynamic spin ice correlations with decreasing temperature in an annealed Ce$_2$Zr$_2$O$_7$ powder sample. (b) The powder-averaged {\it difference} neutron scattering spectra for an annealed single crystal sample of Ce$_2$Zr$_2$O$_7$.  A data set at $T$~=~2~K has been subtracted from that at $T$~=~0.06~K.   (c) A cut along $\vert${\bf Q}$\vert$ through this difference spectra showing that the dominant quasi-elastic signal, integrated in energy between 0 and 0.15~meV, is centred on $\vert${\bf Q}$\vert$~=~$\vert$(001)$\vert$ ($\sim$0.59~\angstrom$^{-1}$) and (d) a comparison of two cuts in energy through the difference spectra shown in (b), with one of these cuts taken with a $\vert${\bf Q}$\vert$-integral centered on $\vert$(001)$\vert$ (0.35 to 0.85~$\angstrom^{-1}$), and one removed from $\vert$(001)$\vert$, integrating between 1.3 and 1.8~$\angstrom^{-1}$. For all these panels, the error bars correspond to one standard deviation.}
\label{DCSdata}
\end{figure}

Additional weak inelastic scattering whose  {\bf Q}-dependence is inconsistent with phonons is also visible in the spectra, for example weak scattering near $\sim$ 100~meV in Fig.1 (b). It is not clear if this weak inelastic scattering is due to the influence of Ce$^{3+}$ or Zr$^{3+}$ in defective sites~\cite{sala2018}, on residual Ce$^{4+}$, or on the possible presence of hybridized phonon - crystal field excitations known as vibronic bound states, as has been recently observed in holmium and terbium pyrochlores~\cite{gaudet2018magneto,ruminy2016crystal}. In any case, this unidentified contribution to the inelastic scattering yields a small fraction of the spectral weight and we conclude the features at 56 and 112~meV are the CEF excitations corresponding to the main Ce$^{3+}$ site. 

The details of the crystal field analysis determining the full set of eigenvalues and eigenfunctions for Ce$^{3+}$ are summarized in Fig.1(c), and further discussed in the SM. The key conclusion is that the ground state Kramers doublet appropriate to Ce$^{3+}$ is well separated from all excited crystal field states (with a gap of $\sim$56~meV), and is composed of pure $m_J$~=~$\pm$ 3/2 states. A large CEF gap is consistent with the high temperature heat capacity of Ce$_2$Zr$_2$O$_7$ measured in ref.~\cite{popa2008re} where no Schottky anomaly is observed between 5 and 300~K. These pure $m_J$~=~$\pm$ 3/2 states have a dipole-octupole character with a dipolar moment whose anisotropy is purely Ising and whose magnitude must be 1.286~$\mu_B$. This result does not originate from a fine-tuning of the CEF parameters, but is instead a property protected by the point-group symmetry of the A-site in the pyrochlore lattice.

\begin{figure*}[tbp]
\linespread{1}
\par
\includegraphics[width=6.5in,height=1.95in]{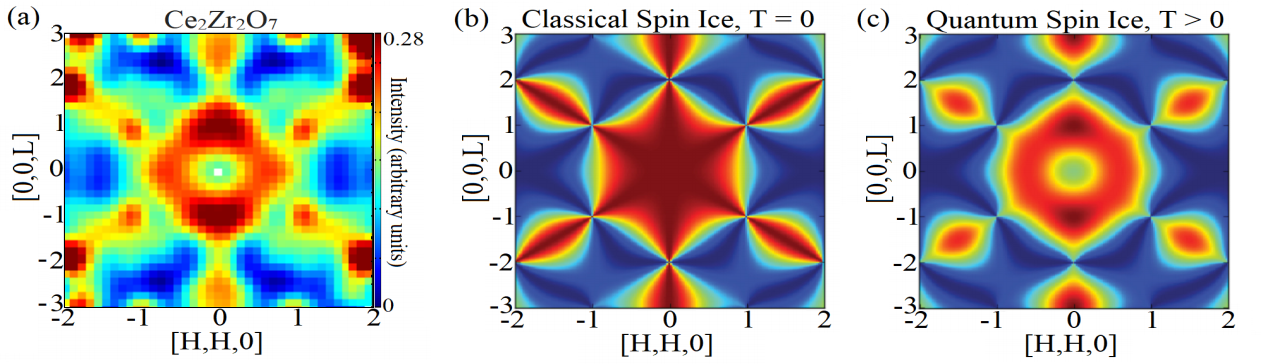}
\par
\caption{Comparison of the measured low energy inelastic neutron scattering from (a) an annealed single crystal sample of Ce$_2$Zr$_2$O$_7$ with the calculated quasi-elastic neutron scattering for (b) the classical near-neighbor spin ice model at $T$~=~0~K and (c) a quantum spin ice at finite T. Data in (a) is the symmetrized difference between inelastic scattering at $T$~=~0.06~K and $T$~=~2~K, integrated between 0 and 0.15 meV. Both (b) and (c) are simulations taken from Benton et al.~\cite{bentonphoton}. The lack of intensity around {\bf Q}~=~(000) and the fact that the ring of diffuse inelastic scattering peaks along (00L) provides evidence for Ce$_2$Zr$_2$O$_7$ displaying a dynamic quantum spin ice state at these low temperatures. Also, the observed diffuse inelastic scattering at {\bf Q}~=~(003) is more pronounced than that at {\bf Q}~=~($\frac{3}{2}\frac{3}{2}\frac{3}{2}$), again consistent with the expectations of quantum spin ice, and not consistent with classical near-neighbor spin ice. Note the extra features centred at the Bragg peak positions such as (111) originate from leakages of the structural Bragg peaks, due to the subtraction of two large intensities.}
\label{DCSdata}
\end{figure*}

Figure 1(a) shows the inverse magnetic susceptibility of a 107~mg powder sample of Ce$_2$Zr$_2$O$_7$ measured with a Quantum Design MPMS magnetometer equipped with a $^{3} $He insert. The dc magnetic susceptibility measurements were made with applied magnetic fields of $\sim$0.001~T, $\sim$0.01~T, and $\sim$0.1~T, and no significant field dependence was observed. The main panel of Fig.1(a) shows the high temperature susceptibility of Ce$_2$Zr$_2$O$_7$ and reveals strong non-linearity. Assuming a dilution of the Ce$^{3+}$ moments by non-magnetic Ce$^{4+}$ ions at the $\sim$8~\% level, the Van-Vleck susceptibility calculated with the CEF Hamiltonian of Ce$_2$Zr$_2$O$_7$ reproduces the high-temperature susceptibility data well and yields an antiferromagnetic Curie constant of -~0.4(2)~K. We expect conventional and unfrustrated AIAO order in Ce$_2$Zr$_2$O$_7$ based on the effective antiferromagnetic interactions and the Ising anisotropy associated  with its magnetism. However, our magnetic susceptibility measurements (inset of Fig.1(a)) as well as both powder and single crystal neutron diffraction experiments show no indication of long range magnetic order down to $T$~=~0.06~K. In particular and as shown in the SM, no new Bragg scattering or enhancement of the Bragg scattering associated with any {\bf k}~=~0 magnetic structure is observed, including at those wave vectors characteristic of the AIAO, $\Gamma_3$ structure. Ce$_2$Zr$_2$O$_7$ therefore remains disordered to $T$~=~0.06~K, our lowest temperature measured. The lack of magnetic order in Ce$_2$Zr$_2$O$_7$ likely indicates a strong easy-axis anisotropy in its exchange Hamiltonian. 

We examined the low-temperature spin dynamics in Ce$_2$Zr$_2$O$_7$ using the low energy DCS neutron chopper spectrometer at NCNR with E$_i$~=~3.27~meV incident neutrons giving an energy resolution of $\sim$0.09~meV at the elastic line. One experiment was performed on a $\sim$~6 gram powder sample and a second one was performed on a $\sim$~5 gram single crystal, which was mounted with its [HHL] plane coincident with the horizontal plane of the spectrometer. Figure 2(a) shows the DCS measurements on our powder, where the integration in  $\vert${\bf Q}$\vert$ is 0.35~$\angstrom^{-1}$ to 0.85~$\angstrom^{-1}$. This integration in momentum transfer $\vert${\bf Q}$\vert$ corresponds to integrating over the  $\vert${\bf Q}$\vert$~=~$\vert$(001)$\vert$ position ($\sim$ 0.59~$\angstrom^{-1}$), where quantum spin ice correlations are expected to be strongest~\cite{bentonphoton}. A build up of inelastic spectral weight below $\sim$~0.4~meV  is observed on decreasing the temperature.

Low energy inelastic neutron scattering from our single crystal is shown in Fig.2(b,c,d) and Fig.3(a). All this data was acquired using the same E$_i$~=~3.27~meV instrument configuration of DCS, and Fig.2(b,c,d) shows {\it powder-averaged} single crystal data. Figure 2(b) shows the full powder-averaged spectrum at $T$~=~0.06~K with a $T~=~2~K$ data set subtracted from it. This result shows enhanced inelastic scattering at low temperature, which peaks up at $\vert${\bf Q}$\vert$ $\sim$~0.59~$\angstrom^{-1}$, that is the magnitude of the {\bf Q}~=~(001) position in reciprocal space. This is explicitly shown via the $\vert${\bf Q}$\vert$-cut of the data presented in Fig.2(c). Importantly, Fig.2(c) shows no enhancement of the low energy inelastic scattering around $\vert${\bf Q}$\vert$~=~0, consistent with expectations for a  U(1) quantum spin ice. Finally, Fig.2(d) shows energy cuts through the full difference spectrum shown in (b), taken by integrating in $\vert${\bf Q}$\vert$ from 0.35 to 0.85~$\angstrom^{-1}$, so around $\vert${\bf Q}$\vert$~=~$\vert$(001)$\vert$, and also well away from $\vert${\bf Q}$\vert$~=~$\vert$(001)$\vert$,  integrating from 1.3 to 1.8~$\angstrom^{-1}$. This clearly shows the quantum spin ice correlations to be dynamic in nature, characterized by an energy less than $\sim$~0.15~meV.

With the energy range of the dynamic quantum spin ice correlations identified, we can look explicitly at this scattering from the single crystal, but now comparing {\bf Q} maps of these correlations to the expectations of both classical, near-neighbor spin ice (without dipolar interactions) and a U(1) quantum spin ice. Fig.3(a) shows $T$~=~0.06~K - $T$~=~2~K data integrated between 0 and 0.15~meV, folded into a single quadrant of the [HHL] map and further symmetrized. The details of this data symmetrization are in the SM.  For reference, a theoretical simulation of the structure factor expected for classical near-neighbor spin ice~\cite{bentonphoton} is shown in Fig.3(b), and that for a U(1) quantum spin ice at low but finite temperature~\cite{bentonphoton} is shown in Fig.3(c). While these theoretical predictions have similarities, the structure factor for U(1) quantum spin ice has minima in intensity near {\bf Q}~=~0, while the intensity of the structure factor is maximal there for classical near-neighbor spin ice.

Clearly, the measured dynamic S({\bf Q}) shows a qualitatively stronger resemblance to the expectations of the symmetry enriched U(1) quantum spin ice~\cite{hermele2004pyrochlore,savary2012coulombic,savary2013spin,bentonphoton}. The quantum spin ice ground-state is one of various spin liquids that are supported by a model of well isolated DO CEF doublets on the pyrochlore lattice~\cite{huang2014quantum,li2017}.  A similar dynamic S({\bf Q}) is expected in the case of classical {\it dipolar} spin ice (here {\it dipolar} refers to long range dipolar interactions between magnetic dipoles, as opposed to quadrupoles, octupoles etc), which also shows the suppression of diffuse scattering near $\vert${\bf Q}$\vert$~=~0~\cite{den2000dipolar,bramwell2001dip}. Although a definitive conclusion can only be reached once a full spin Hamiltonian is parametrized, the Ce$^{3+}$ ions in Ce$_2$Zr$_2$O$_7$ have a moment of 1.286~$\mu_B$.  These moments are roughly a factor of 8 smaller than those associated with Ho$^{3+}$ or Dy$^{3+}$ in the classical dipolar spin ices Ho$_2$Ti$_2$O$_7$ and Dy$_2$Ti$_2$O$_7$. The resulting long range dipole terms are expected to be $\sim$~64 times weaker in Ce$_2$Zr$_2$O$_7$, making such a scenario unlikely. This suggests the spin-ice correlations in Ce$_2$Zr$_2$O$_7$  originate from quantum effects. An octupolar ordered state is also consistent with the lack of magnetic dipole order in Ce$_2$Zr$_2$O$_7$. However, the neutron scattering spectra associated with such an octupolar ordered phase has yet to be calculated, thus we cannot compare our data in Fig.3(a) to it.

The effect of disorder in Ce$_2$Zr$_2$O$_7$ is still an open question as we are aware that our single crystals have some low levels of oxidation. Furthermore, stuffing~\cite{ross2012lightly,koohpayeh2014synthesis,baroudi2015,arpino2017impact} (site-mixing) is expecting to be important in Ce$_2$Zr$_2$O$_7$, because both undesired Ce$^{4+}$ and Zr$^{3+}$ ions are chemically stable. It is known that small amounts of disorder can have a drastic impact on the physical properties of frustrated pyrochlore magnets~\cite{ross2012lightly,koohpayeh2014synthesis,arpino2017impact}. It will then be important to further optimize the growth procedure and annealing techniques of Ce$_2$Zr$_2$O$_7$. However, we believe that our inelastic neutron scattering results rule out the scenario of a sensitive AIAO order. Indeed, the conventional impact of quenched disorder on a pyrochlore antiferromagnet would be spin glass physics with diffuse scattering peaked for {\bf Q}'s corresponding to the Bragg positions of the AIAO state. Here, we observe strong diffuse scattering at {\bf Q}~=~(001), which is not only strictly zero for an AIAO state, but also forbidden for all {\bf k}~=~0 long-range ordered magnetic structures allowed by symmetry of the pyrochlore lattice. We thus conclude that our work demonstrates Ce$_2$Zr$_2$O$_7$ to be one of a very few candidates for quantum spin ice physics. Other candidates for quantum spin ice physics are based on Pr$^{3+}$ and Tb$^{3+}$ pyrochlores~\cite{zhou2008dynamic,fritsch2013anti,kimura2013quantum,sibille2016candidate,sibille2018experimental}. However, in contrast to Pr$^{3+}$ and Tb$^{3+}$, Ce$^{3+}$ is a Kramers ion and its magnetism is thus further protected against disorder, which in and of itself, can drive a spin liquid state for non-Kramers doublets~\cite{savary2017disorder,wen2017disordered,martin2017disorder,benton2018inst}. Furthermore, Tb$^{3+}$ and Pr$^{3+}$ pyrochlores display low lying CEF field states, which complicate their theoretical understanding due to multipolar interactions~\cite{molavian2007dynamically,takatsu2016quadrupole,petit2016antiferroquadrupolar}. For all these reasons, the cerium pyrochlores are an excellent theoretical and experimental template to investigate quantum spin ice physics.  

\begin{acknowledgments}
We acknowledge useful conversations with Allen Scheie and Alannah Hallas. We greatly appreciate the technical support from Alan Ye and Yegor Vekhov at the NIST Center for Neutron Research. This work was supported by the Natural Sciences and Engineering Research Council of Canada (NSERC). A portion of this research used resources at the High Flux Isotope Reactor and Spallation Neutron Source, a DOE Office of Science User Facility operated by the Oak Ridge National Laboratory. We also acknowledge the support of the National Institute of Standards and Technology, U.S. Department of Commerce. Certain commercial equipment, instruments, or materials (or suppliers, or software, etc) are identified in this paper to foster understanding. Such identification does not imply recommendation or endorsement by the National Institute of Standards and Technology, nor does it imply that the materials or equipment identified are necessarily the best available for the purpose. 
\end{acknowledgments}

\bibliography{CeZrO_Bib}

\section{Supplemental Material:}

\subsection{Powder synthesis and Single crystal growth}
The polycrystalline and powder samples of Ce$_2$Zr$_2$O$_{7}$ used in this work were first prepared by arc melting stoichiometric amounts of CeO$_2$, Zr, and ZrO$_2$ in an argon atmosphere, followed by a regrinding and firing at 1000$^\circ$C for two days in flowing hydrogen. Prior to any measurements, the powder samples were further annealed in flowing hydrogen at 1000$^\circ$C for several hours. An x-ray refinement against the $Fd\overline{3}m$ space group is shown in Fig.S1(a) for one of the Ce$_2$Zr$_2$O$_7$ powder samples, synthesized and annealed using this protocol. Previous studies have characterized the level of oxidation of Ce$_2$Zr$_2$O$_{7+ \delta}$ powder samples using the value of the lattice parameter, $a$, refined from x-ray diffraction measurements. For example, the values $a$~$\sim$~10.735~$\angstrom$ has been refined for  $\delta$~=~0 and $a$~$\sim$~10.66~$\angstrom$ has been refined for  $\delta$~=~0.5. A linear relationship has been observed between these limits~\cite{otobe2005,thomson1999,thomson1996}. Our materials synthesis and powder x-ray diffraction is consistent with these previous works, and, as shown in Fig.S1(a), our stoichiometric powder sample has a lattice parameter of $a$~=~10.735(5)~$\angstrom$. As we will discuss, we observe the oxidation of the powder sample left exposed to air to begin to occur on the order of minutes. There is an obvious  colour change which occurs between the stoichiometric (annealed) and oxidized samples, wherein a light green/yellow sample is obtained for the $\delta$~=~0 annealed sample, and a black powder is obtained for the $\delta$~=~0.5 oxidized sample. 

\begin{figure}[tbp]
\linespread{1}
\par
\includegraphics[width=3.4in]{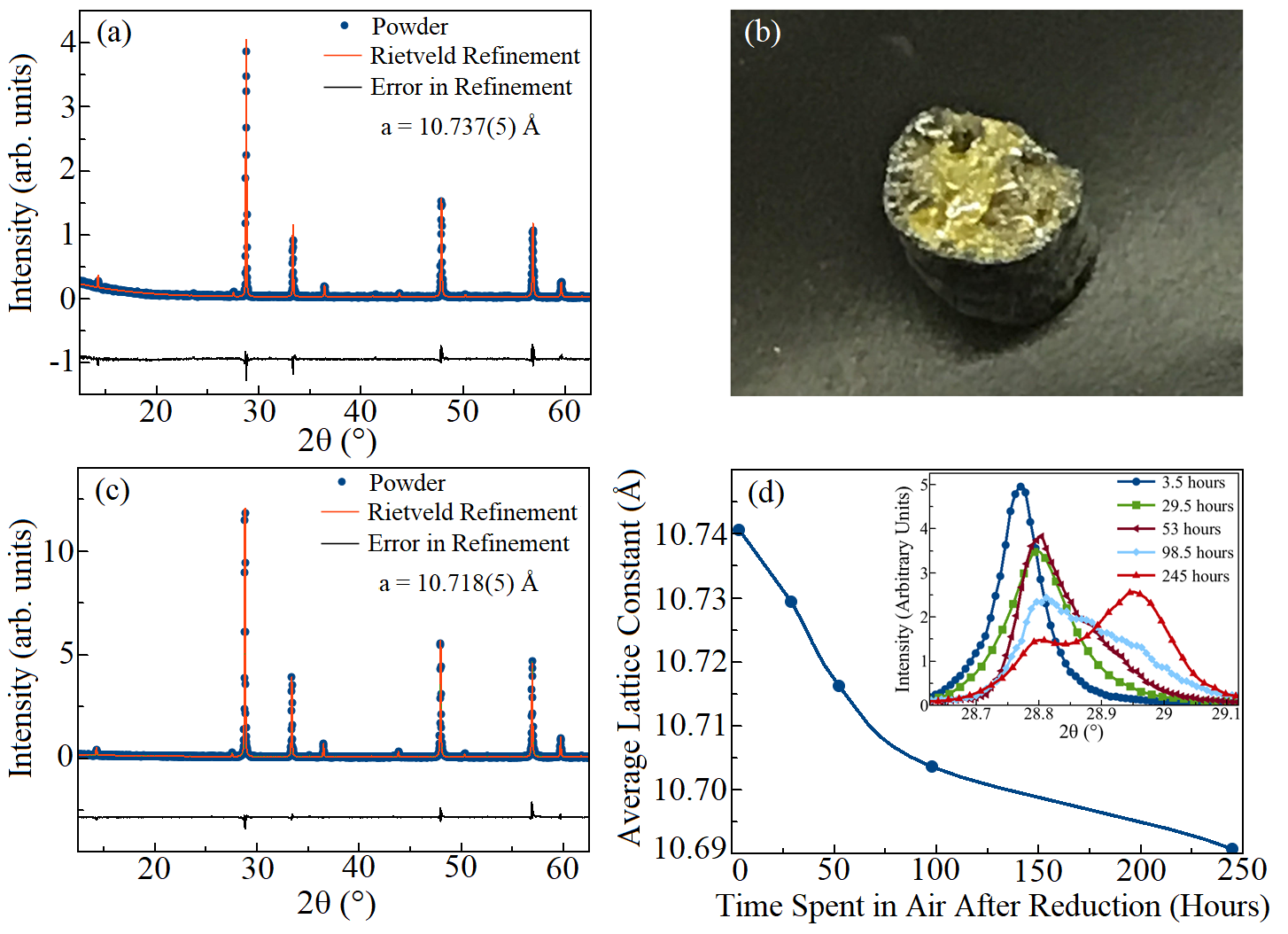}
\par
\caption{(a) Powder x-ray refinement of a typical powder sample of Ce$_2$Zr$_2$O$_7$ synthesized for this work. A lattice parameter of 10.735(5)~$\angstrom$ has been refined. (b) A photograph of an annealed single crystal sample of Ce$_2$Zr$_2$O$_7$, broken with an exposed surface,  and showing a bright yellow color on the inside and a thin black oxidized surface on the outside. (c) Powder x-ray refinement of a crushed single crystal of Ce$_2$Zr$_2$O$_7$ against the pyrochlore structure. (d) The time dependence of the lattice parameter is shown for a polycrystalline sample of Ce$_2$Zr$_2$O$_7$, annealed in hydrogen at 1200$^\circ$C for 6 hours, and then left exposed to air. The exposure to air produces oxidation of the sample over time and a decreasing lattice parameter. The inset shows typical x-ray diffraction scans collected at different times of exposure to air, following the annealing protocol.}
\label{SAMPLE}
\end{figure}

\begin{figure*}[tbp]
\linespread{1}
\par
\includegraphics[width=6.9in]{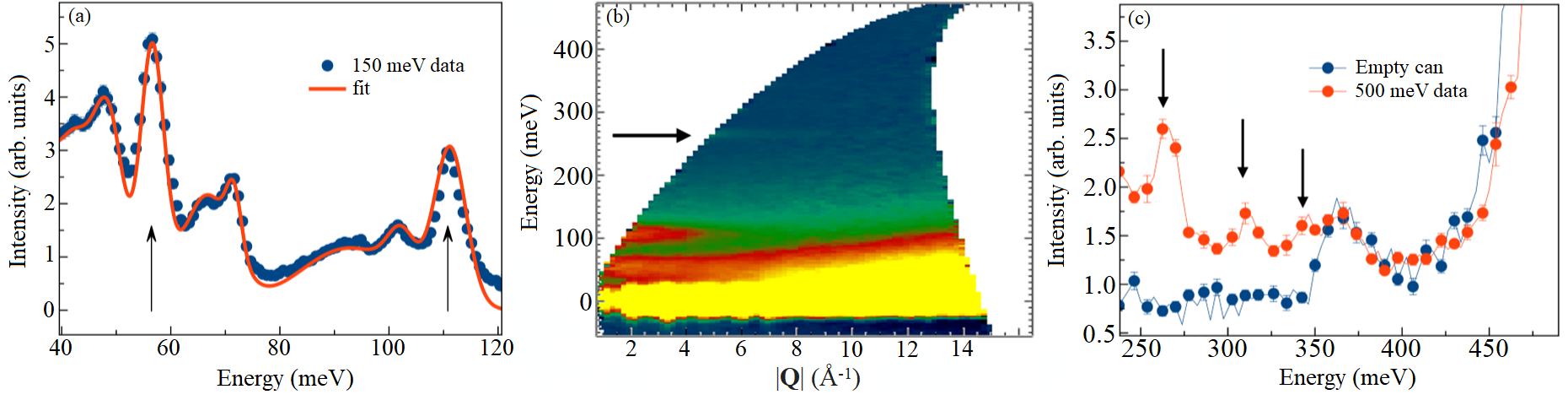}
\par
\caption{(a)  Inelastic neutron scattering spectra obtained for an incident energy of 150~meV for Ce$_2$Zr$_2$O$_7$, and integrated between $\vert${\bf Q}$\vert$~=~4.5 and 5.5~\angstrom$^{-1}$ (this is a cut of the data shown in Fig.1(a) of the main manuscript). The two black arrows indicate the energy position of the CEF transitions originating from the main Ce$^{3+}$ site. (b) High energy inelastic neutron scattering spectra of Ce$_2$Zr$_2$O$_7$ obtained with an incident energy of 500~meV. The arrow highlights the relatively weaker magnetic excitations near 275~meV. c) A constant energy cut of the data shown in b), integrated in $\vert${\bf Q}$\vert$ between 7 and 10~\angstrom$^{-1}$, is shown, along with data from the empty sample can. We identify three CEF transitions from the ground state to the highest $J$ manifold at $\sim$ 270 meV, 310 meV and 340 meV. These energies are sufficiently high, such that no optic phonons are expected to be nearby in energy.}
\label{CEF}
\end{figure*}

Single crystals of Ce$_2$Zr$_2$O$_7$ were obtained using optical floating zone growth, with Xe lamps, from polycrystalline feed stock. The powder rods were first prepared through a solid state reaction using high purity materials. Stoichiometric mixtures of Ce$_2$O$_3$ (99.995$\%$) and ZrN (99.5$\%$) were mixed in a ball mill and pressed into rods. The rods were heated in air to 900$^\circ$C in a covered alumina crucible for 5~h. The solidified rods were then re-ground in a ball mill, repressed into rods 8~cm and 6~cm in length to be used as the feed and seed rods for the optical floating zone growth. These rods were then heated to 1550$^\circ$C for 3~h in an atmosphere containing a ratio of 95/5 argon to hydrogen. During the actual optical floating zone growth, we employed a growth rate of 2.5 mm/hour while counter-rotating feed and seed rods at 15 rpm in an argon atmosphere containing 5$\%$ hydrogen. Further annealing of the single crystals for 36 hours at 1200$^\circ$C was also performed prior to all experiments. A photograph of a typical single crystal of Ce$_2$Zr$_2$O$_7$ obtained through this protocol is shown in Fig.S1(b).  A broken surface is shown and one can see that the inside of the single crystal piece is bright yellow while the surface exposed to air is black, indicating some remaining oxidation at the surface after the annealing process. Refinement of a yellow single crystal piece from the unexposed inner layer of the crystal is shown in Fig.1(c). Such x-ray diffraction measurements on crushed single crystals give a lattice parameter refinement of $a$ = 10.718(5)~$\angstrom$ for the bulk inner layer of the crystal, as shown in Fig.1(c), and this can be use to estimate the oxidation state of our annealed single crystal sample, Ce$_2$Zr$_2$O$_{7 + \delta}$, to be $\delta\sim$ 0.1. 

Finally, we characterized the time dependence of Ce$_2$Zr$_2$O$_{7 + \delta}$ by collecting several powder diffraction patterns from a broken, ceramic rod of material, for differing exposure times in air following a 1200$^\circ$C, 6 hour hydrogen annealing protocol. The lattice parameter extracted from x-ray diffraction measurements is shown as a function of time in Fig.S1(d) for this polycrystalline sample, whose oxidation rate should be between the fast rate of the powder samples of Ce$_2$Zr$_2$O$_{7+\delta}$ and slow rate of the relatively-well behaved single crystal samples of Ce$_2$Zr$_2$O$_{7+\delta}$. The inset shows x-ray diffraction scans of the {\bf Q}~=~(222) Bragg peak as a function of scattering angle and exposure time. Not only do the Bragg peaks shift higher in 2$\theta$ with exposure (sample oxidation), they also significantly broaden and eventually split into multiple distinguishable peaks. This broadening and splitting corresponds to a distribution of oxidation in the sample, and to the eventual formation of separate majority and minority phases with different levels of oxidation, such as the bulk inner and thin outer layers of differing color in the case of single crystal Ce$_2$Zr$_2$O$_{7 + \delta}$. The x-ray intensity at the {\bf Q}~=~(222) Bragg position for each data set was fit to Lorentzian peak shapes according to the number of peaks present around that Bragg position. The average lattice constant for each data set with multiple distinguishable peaks ($>$ 29.5 hours) was determined by using a weighted average of the lattice constants determined from these peaks, with the integrated intensity of the peaks used as the weights for this average. Figure S1(d) should then give the time dependence of the volume-averaged lattice constant in this polycrystalline sample.

\subsection{Determination of the CEF Hamiltonian}

As discussed in the main manuscript, the CEF eigenstates for Ce$^{3+}$ in Ce$_2$Zr$_2$O$_7$ were expressed in the $\ket{J=5/2,m_J}$ manifold and fitted using the low energy CEF transitions observed in the E$_i$~=~150~meV neutron scattering spectra. We refined a CEF Hamiltonian for Ce$_2$Zr$_2$O$_7$ by constraining the relative scattered intensities and energies of the CEF transitions observed in our E$_i$~=~150~meV spectra (Fig.1(a) of the main letter). The following CEF Hamiltonian was used in this work:
\begin{eqnarray}
\mathcal{H}_{CEF}= B^0_2\hat{O}^0_2 + B^0_4\hat{O}^0_4 +  B^3_4\hat{O}^3_4
\label{eq: HCEF}
\end{eqnarray}

The protocol and the expression of the CEF Hamiltonian in terms of Stevens operators used in this work is identical to ref.~\cite{gaudet2015neutron,gaudet2018effect} and we refer the reader to these works for further details. It is worthwhile mentioning that the CEF Hamiltonian appropriate for the R$^{3+}$(D$_{3d}$) site of the pyrochlore lattice includes six Stevens operators (B$_2^0$, B$_4^0$, B$_4^3$, B$_6^0$, B$_6^3$, B$_6^6$), however the three B$_6^n$ terms are zero for Ce$^{3+}$.

The B$_2^0$, B$_4^0$ and B$_4^3$ terms of the CEF Hamiltonian were refined using the observed energies and scattered neutron intensities of the magnetic excitations near 58 and 110~meV. It was possible to extract this information using a constant $\vert${\bf Q}$\vert$-cut from the 150~meV data.  We obtain this by integrating the 150~meV spectra (Fig.1(a) of the main letter) between $\vert${\bf Q}$\vert$~=~4.5 and 5.5~\angstrom$^{-1}$. The resulting integration is shown in Fig.S2(a). The two CEF transitions, so-identified by their $\vert${\bf Q}$\vert$-dependence, are indicated by the black arrows in Fig.S2(a), while other features observed in Fig.S2(a) are likely optical phonon excitations. The CEF transitions as well as the phonon excitations were fitted using a Lorentzian function in energy. The resulting fit of the 150~meV data set is plotted in Fig.S2(a). From this fit, we determined that the energies of the CEF levels within the spin-orbit $J$-manifold are E$_1$~=~55.9(1) and E$_2$~=~110.5(1)~meV. The relative intensity of the CEF transitions between the CEF ground state to the first (I$_1$) and the second (I$_2$) excited state is found to be I$_1$/I$_2$~=~1.2(1). Using these three constraints (E$_1$, E$_2$, I$_1$/I$_2$), we refined B$_2^0$~=~0.835~meV, B$_4^0$~=~0.299~meV and B$_4^3$~=~2.875~meV. Comparison between the calculation and data is shown in Table.S1. The calculated eigenstates and eigenfunctions using this CEF Hamiltonian are reported in Table.S2.

In the main letter, we also reported a weaker magnetic excitation near 100~meV, so just below E$_2$. It is possible that magneto-elastic coupling between the CEF ground state and the 2nd excited CEF state (E$_2$) is sufficiently strong to drive a vibronic bound state, which would effectively split the single-ion CEF excitation at E$_2$ into two different excitations~\cite{gaudet2018magneto}. Here, these two excitations would correspond to the one at 110~meV and the one at 100~meV. Thus, within the scenario of a vibronic bound state as the origin of the inelastic scattering at 100~meV, we underestimated I$_2$.  We can compensate for this by adding the integrated intensity of the 100~meV transition. By doing so, a new optimization of the CEF Hamiltonian could be refined, leading to B$_2^0$~=~0.455~meV, B$_4^0$~=~0.295~meV and B$_4^3$~=~2.582~meV. Table.S1 also shows the comparison between the data and the calculation within such a scenario. This scenario in which a vibronic bound state is responsible for the weak inelastic intensity near 100 meV also leads to a CEF ground state that is a pure m$_J$~=~$\pm$3/2 state.  Thus the conclusion that the CEF ground state doublet for Ce$_2$Zr$_2$O$_7$ is a pure  m$_J$~=~$\pm$3/2 doublet with dipolar octupolar character is robust. 

We also present the E$_i$~=~500~meV inelastic neutron scattering spectra of Ce$_2$Zr$_2$O$_7$ in Fig.S2(b). This inelastic scattering data shows a clear magnetic feature near 275~meV that corresponds to a transition originating from the CEF ground state to the highest $J$-manifold ($J$~=~7/2). Two additional, albeit weaker, excitations can be identified by taking appropriate $\vert${\bf Q}$\vert$ cuts of this data and comparing to the measured empty can, background scattering. This is shown in Fig.S2(c), where a $\vert${\bf Q}$\vert$ integration from $\vert${\bf Q}$\vert$~=~7 to 10~\angstrom$^{-1}$ is performed. This reveals three CEF transitions $\sim$~260, 310 and possibly 340~meV. The location in energy of these highest $J$~=~7/2 states is consistent with estimates of the spin-orbit coupling strength ($\lambda$) for Ce$^{3+}$~\cite{Freeman1962,Blume1964}. 

Finally, the determination of our CEF Hamiltonian appropriate for Ce$_2$Zr$_2$O$_7$ was further validated via calculation of its Van-Vleck susceptibility that can be compared with the temperature dependence of its measured magnetic susceptibility. The following Van-Vleck susceptibility term (ref.\cite{VanVleck})  was computed for a powder sample:
\begin{eqnarray}
\chi_{CEF} = \frac{N_A g^2_J \mu^2_B X}{k_B Z}\sum_\alpha(\sum_n\frac{|\bra{n}J_\alpha\ket{n}|^2e^{E_n/T}}{T}+\\
\nonumber
\sum_n\sum_{m\neq n}\frac{|\bra{m}J_\alpha\ket{n}|^2(e^{-E_n/T}-e^{-E_m/T})}{E_m-E_n})
\end{eqnarray}
where $\alpha = x,y,z$, N$_A$ is the Avogadro constant, $g_J$ is the Land\'e $g$-factor, $k_B$ is the Boltzman constant, $\mu_B$ is the Bohr magneton and $Z=\sum_n e^{-E_n/T}$ is the partition function. The factor $X$ was used to parametrize the dilution of the Ce$^{3+}$ ions into non-magnetic Ce$^{4+}$ and was refined to 0.92(2), which is in good agreement with our estimate of the oxidation level in our samples using the refined lattice parameters  (see section 1 of the SM).  
\begin{table}[]
\caption{Comparison between the observed (Obs$_1$,Obs$_2$) and calculated (Calc$_1$,Calc$_2$) energies and intensities of the CEF excitations in Ce$_2$Zr$_2$O$_7$. Scenario 1 refers to the case where no vibronic bound state is present, while Scenario 2 is the case where weak inelastic scattering near 100 meV is identified as arising due to a vibronic bound state.}
\begin{tabular}{|c|c|c|c|c|}
\hline
                                                                           & \textbf{Obs$_1$} & \textbf{Calc$_1$} & \textbf{Obs$_2$} & \textbf{Calc$_2$} \\ \hline
\textbf{\begin{tabular}[c]{@{}c@{}}E$_1$\\ (meV)\end{tabular}}             & 55.9(2)       & 55.92          & 55.9(1)       & 56.06          \\ \hline
\textbf{\begin{tabular}[c]{@{}c@{}}E$_2$\\ (meV)\end{tabular}}             & 110.5(3)      & 110.55         & 106.1(4)      & 105.97         \\ \hline
\textbf{\begin{tabular}[c]{@{}c@{}}I$_1$/I$_2$\\ (arb.units)\end{tabular}} & 1.2(1)        & 0.98           & 0.99(15)      & 0.92           \\ \hline
\end{tabular}
\end{table}

\begin{table}[]
\caption{Eigenstates and eigenfunctions of the spin-orbit ground state manifold written within the $\ket{J=5/2,m_J}$ basis. These eigenfunctions correspond to the scenario of no vibronic bound-state. A m$_J$~=~$\pm$3/2 CEF ground state is stabilized for both scenarios (without and with a vibronic bound state).}
\begin{tabular}{|c|c|c|c|c|c|c|}
\hline
\textbf{E(meV)} & \textbf{-5/2} & \textbf{-3/2} & \textbf{-1/2} & 1/2    & \textbf{3/2} & \textbf{5/2} \\ \hline
\textbf{E$_1$}  & 0             & 1             & 0             & 0      & 0            & 0            \\ \hline
\textbf{E$_2$}  & 0             & 0             & 0             & 0      & 1            & 0            \\ \hline
\textbf{E$_3$}  & 0.725         & 0             & 0             & 0.688  & 0            & 0            \\ \hline
\textbf{E$_4$}  & 0             & 0             & -0.688        & 0      & 0            & 0.725        \\ \hline
\textbf{E$_5$}  & 0             & 0             & -0.725        & 0      & 0            & -0.688       \\ \hline
\textbf{E$_6$}  & 0.688         & 0             & 0             & -0.725 & 0            & 0            \\ \hline
\end{tabular}
\end{table}

\subsection{Details on the lack of magnetic order in Ce$_2$Zr$_2$O$_7$}

An orange cryostat with a dilution refrigerator insert was used for both of our DCS experiments. The powder sample was wrapped in of a copper foil inside a copper can sealed under 10~atm of He. For the single crystal experiment, the sample was aligned and mounted on a copper mount. For the data collection of our single crystal DCS experiment, the single crystal was rotated for a total of 270$^\circ$ with steps of 0.5~$^\circ$ and we counted for 5 minutes per angle at $T$~=~0.06~K and 3 minutes per angle at $T$~=~2~K. The data was analyzed and plotted using DAVE~\cite{Dave}.

We demonstrate the evidence for no magnetic order in Ce$_2$Zr$_2$O$_7$ to temperatures as low as $T$~=~0.06~K using our single crystal DCS experiment. We do this by noting that no new Bragg peaks are observed at $T$~=~0.06~K compared with $T$~=~2~K, and by explicitly isolating the elastic scattering for each {\bf k}~=~0 Bragg peak accessible in our experiment. This elastic scattering is shown in Fig.S3 at both $T$~=~0.06~K and $T$~=~2~K for the ${\bf Q}$ = (220),(113),(111),(222) and (002) positions of reciprocal space, with $E_i$~=~3.27~meV incident neutrons and energy integration in the range from -0.1 to 0.1~meV. None of these selected {\bf k}~=~0 positions in reciprocal space show any significant changes in intensity with temperature as can be seen by the subtraction of the $T$~=~2~K data set from the $T$~=~0.06~K data set in each plot of Fig.S3. The elastic cuts through the ${\bf Q}$~=~(220),(113) and (111) positions were taken along the (HH0) direction with integration in (00L) from L~=~-0.1 to 0.1 r.l.u., 2.9 to 3.1 r.l.u., and 0.9 to 1.1 r.l.u. respectively and the elastic cuts through the ${\bf Q}$~=~(222) and (002) positions were taken along the (00L) direction with integration in (HH0) from H~=~1.9 to 2.1 r.l.u. and -0.1 to 0.1 r.l.u. respectively. Figure S3(f) shows a table qualitatively outlining which of these positions is expected to show magnetic intensity for each {\bf k}~=~0 magnetic structure permitted by the pyrochlore lattice~\cite{hallas2016xy}. A green check mark indicates the presence of magnetic intensity at that location due to the corresponding ordered structure, while the red symbol indicates that the corresponding structure results in no magnetic intensity at that location. For example, the $\Gamma_3$ structure corresponding to AIAO order generates magnetic intensity at the (220) and (113) positions only. As can be seen from this Table, the fact that we have measured no magnetic intensity at each of the (220), (111), (113), (222) and (002) positions, signifies a lack of ${\bf k}$~=~0 magnetic order in Ce$_2$Zr$_2$O$_7$ down to $T$~=~0.06~K. Furthermore, the whole elastic ${\bf Q}$ map within the (HHL) plane can be plotted for 0.06~K and this is shown in Fig.S3(g). Only the expected structural Bragg peaks are visible in Fig.S3(g), which confirms the lack of ${\bf k}$=0 magnetic order. 

We note that our powder and single crystal DCS experiments that both show no magnetic order in Ce$_2$Zr$_2$O$_7$ are further corroborated with cold neutron triple-axis experiment using the SPINS instrument, also at NIST. For this experiment, the sample was also mounted on a copper mount and aligned within the (HHL) plane. An orange cryostat with an He3 insert was used for this experiment. Figure S4 shows $\theta$-2$\theta$ scans for several Bragg peaks at both 5.2~K and 0.3~K collected with neutron incident energies of 5~meV. It is clear from the temperature difference data that our subtraction of both temperature data sets that our triple-axis experiment also does not support the existence of ${\bf k}$~=~0 magnetic order in Ce$_2$Zr$_2$O$_7$ down to 0.3~K. 

Finally, low temperature magnetic susceptibility measurements of  Ce$_2$Zr$_2$O$_7$ were performed in several fields with both field cooled (FC) and zero field cooled (ZFC) protocol. As seen in Fig.S5, our low temperature susceptibility measurements do not reveal any sign of magnetic ordering or spin freezing down to $\sim$~0.5~K. Furthermore, the general behaviour of the susceptibility is not impacted by the field value used for these measurements. 

\subsection{Details of the symmetrization process in our low energy inelastic neutron scattering (DCS) experiment on a single crystal}

In this section, we outline the symmetrization process used to average the {\bf Q}-dependence of the inelastic temperature difference data from an annealed single crystal sample of Ce$_2$Zr$_2$O$_7$ using the DCS instrument, shown in Fig.3 of the main manuscript. Figure S6(a) displays the subtraction of a $T$~=~2~K data set from a $T$~=~0.06~K data set, integrated in energy over the range from 0 to 0.15~meV , with $E_i$~=~3.27~meV incident neutrons. This is the unsymmetrized data that was used to produce the symmetrized data shown in Fig.3(a) of the main letter. The intermediate steps in this symmetric averaging are shown in Fig.S6(b) and Fig.S6(c) where the data has been folded upon itself and averaged with respect to the (00L) and (HH0) lines of symmetry. The fully symmetrized data is displayed in Fig.S6(d) to further illustrate the process. Only one quadrant of this data set is truly independent; the other three quadrants are required to impose the symmetry of the reciprocal lattice.  No other constraints are imposed on the data. This data set corresponds to symmetric averaging with respect to both the (00L) and (HH0) lines of symmetry. This symmetrization helps to bring out relatively weak signals, as such signals are measured at more than one equivalent {\bf Q} position, and averaged over.  Non-symmetric features in the scattering, resulting from, for example self-absorption, are also averaged over and hence diminished in the process.

\begin{figure*}[tbp]
\linespread{1}
\par
\includegraphics[width=6.7in]{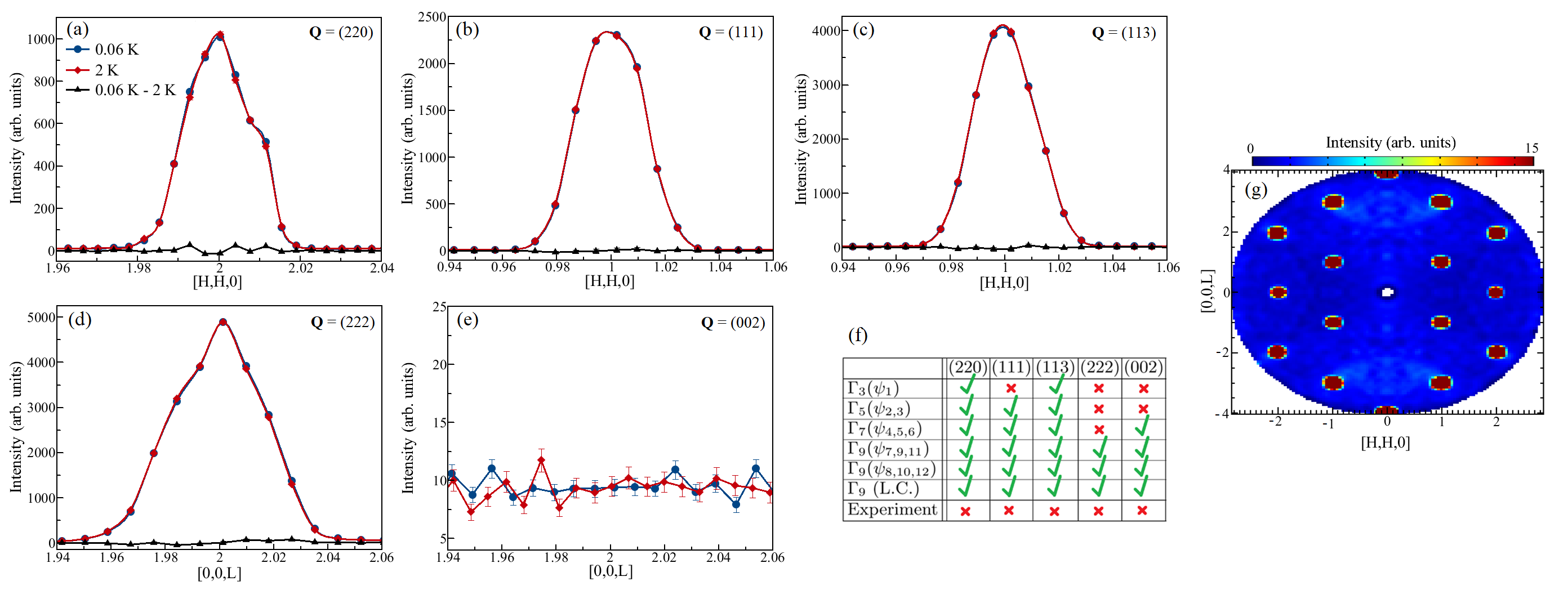}
\par
\caption{Elastic neutron scattering data from an annealed single crystal of Ce$_2$Zr$_2$O$_7$ collected using a time-of-flight neutron spectrometer. Elastic cuts in reciprocal space are shown at $T$~=~0.06~K (blue) and $T$~=~2~K (red) as well as the $T$~=~2~K data set subtracted from the $T$~=~0.06~K data set (black). This elastic data has been integrated from -0.1~meV to 0.1~meV in energy. No significant changes in intensity were measured at the (a) (220), (b) (111), (c) (113), (d) (222), or (e) (002) positions of reciprocal space. (f) A table qualitatively outlining which of these {\bf Q}  positions is expected to show magnetic intensity for the different {\bf k}~=~0 magnetic structures permitted by the pyrochlore lattice~\cite{hallas2016xy}. The green check marks indicate the presence of magnetic intensity at that location due to the corresponding ordered structure while the red symbols indicate that the structure results in no magnetic intensity at that location. As can be seen from this table, the fact that the measured positions showed no systematic change of intensity with temperature indicates a lack of  {\bf k}~=~0 magnetic order down to $T$~=~0.06~K. (g) A  $T$~=~0.06~K map showing the elastic scattering for momentum transfer within the (HHL) plane, which rules out the possible presence of magnetic Bragg peaks with non ${\bf k}$~=~0 ordering vector.} 
\end{figure*}
 
\begin{figure*}[tbp]
\linespread{1}
\par
\includegraphics[width=6.7in]{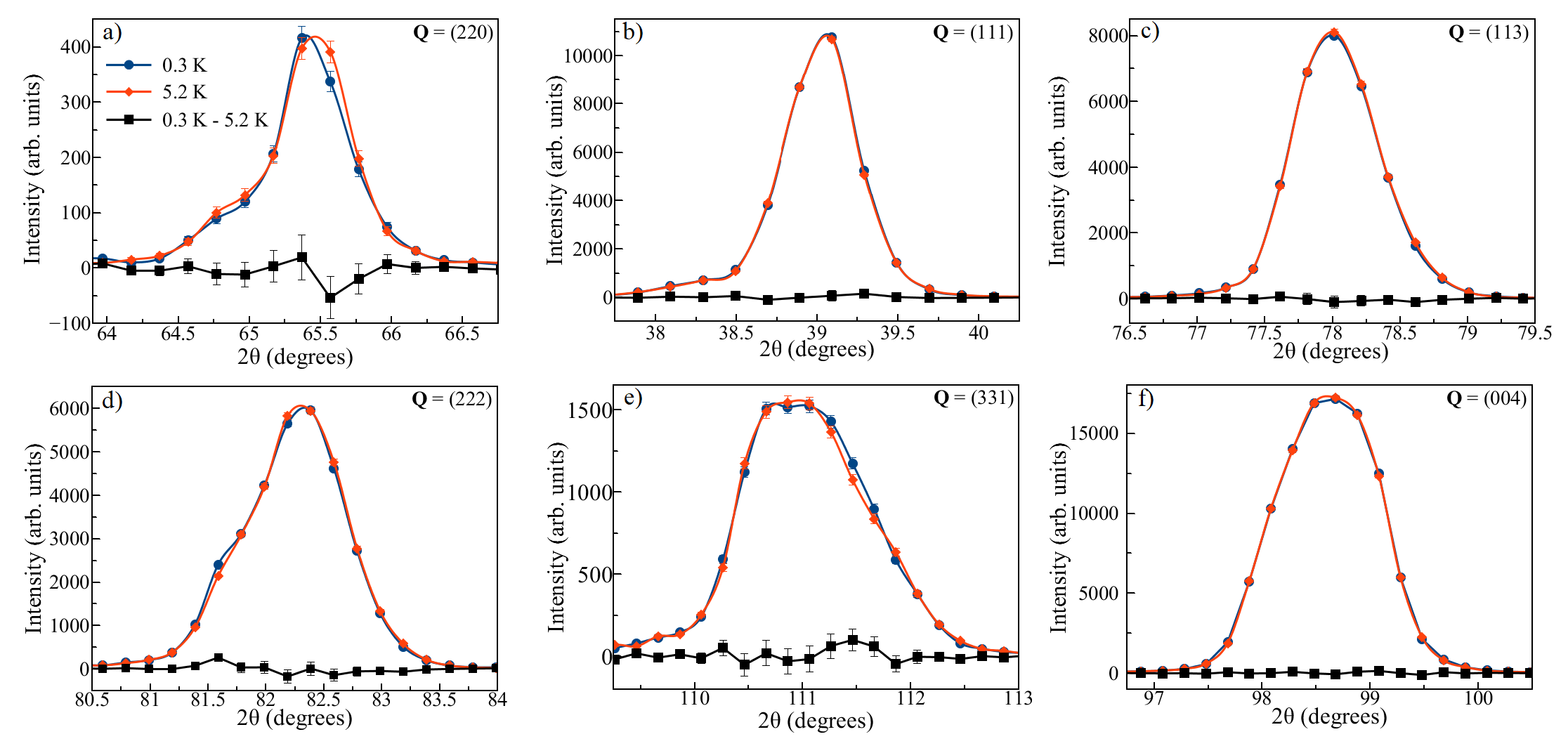}
\par
\caption{$\theta$-2$\theta$ scans for  (a)  {\bf Q} = (220), (b) {\bf Q} = (111), (c) {\bf Q} = (113), (d) { \bf Q} = (222), (e) {\bf Q} = (004) and (f) {\bf Q} = (331) collected on an annealed single crystal of Ce$_2$Zr$_2$O$_7$.  These measurements were performed using a neutron triple-axis instrument at a temperature of 5.2~K (red) and 0.3~K (blue). The difference plots between the two data sets at different temperatures are also shown in each panels (black line) and further confirmed that our single crystal of Ce$_2$Zr$_2$O$_7$ does not magnetically order at low temperature.}
\end{figure*}

\begin{figure*}[tbp]
\linespread{1}
\par
\includegraphics[width=3in]{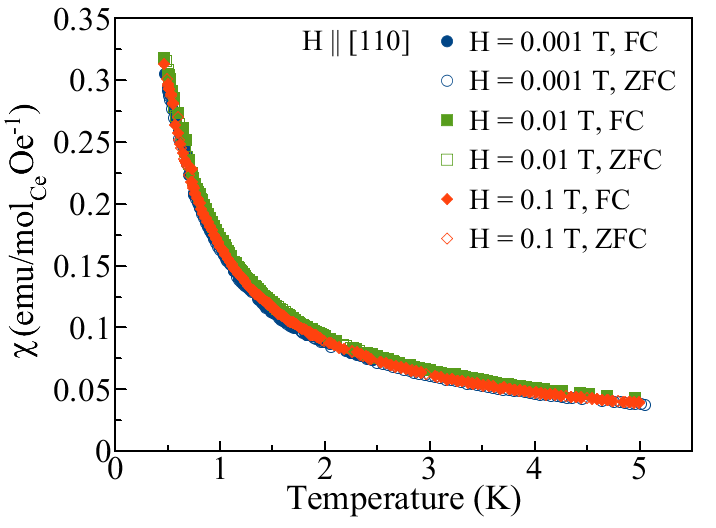}
\par
\caption{The DC magnetic susceptibility measurements on annealed single crystals of Ce$_2$Zr$_2$O$_7$ collected in fields of 0.1~T, 0.01~T and 0.001~T for both field cooled (FC) and zero field cooled (ZFC) protocol. The absence of features in any of these measurements indicates that our single crystal of Ce$_2$Zr$_2$O$_7$ does not undergo a transition to long range magnetic order, or spin freezing down to T~$\sim$~0.5~K.}
\end{figure*}

\begin{figure*}[tbp]
\linespread{1}
\par
\includegraphics[width=5.0in]{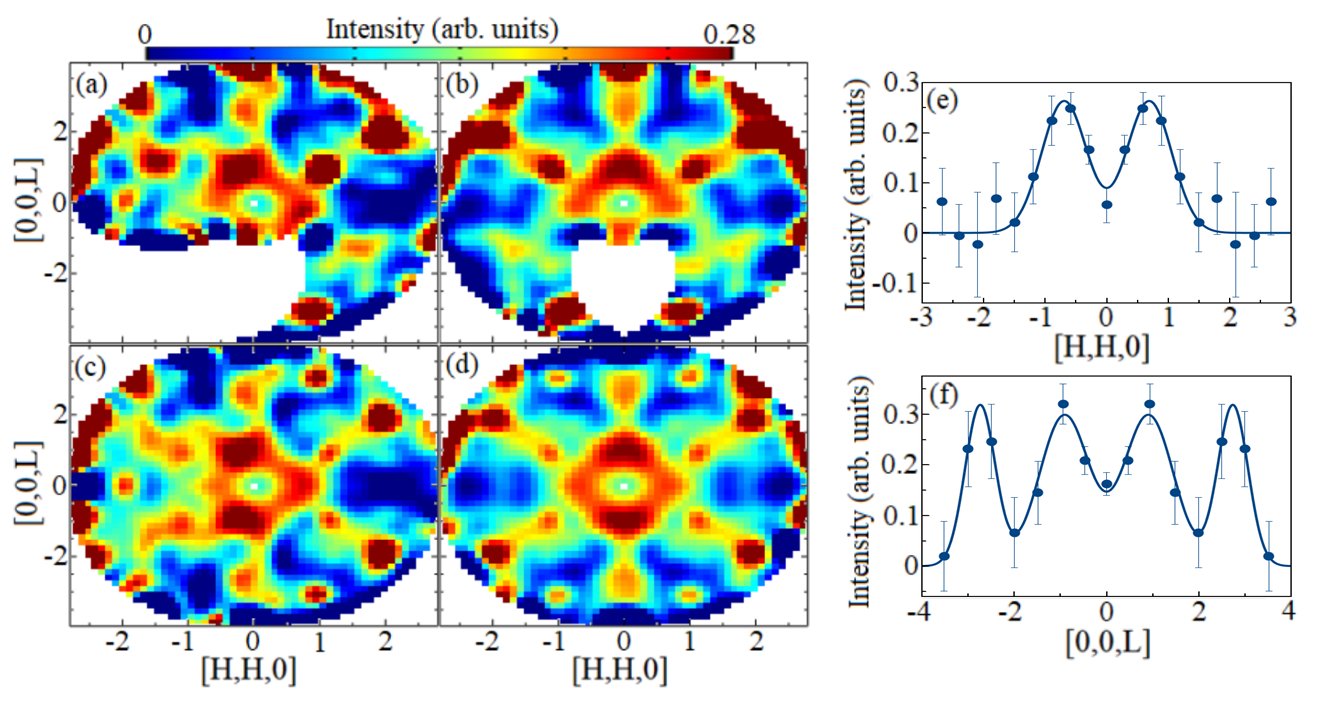}
\par
\caption{The low energy inelastic neutron scattering difference from an annealed single crystal sample of Ce$_2$Zr$_2$O$_7$. This displays the subtraction of a $T$~=~2~K data set from a $T$~=~0.06~K difference data set integrated in energy over the range from 0 to 0.15~meV , with $E_i$~=~3.2~meV incident neutrons. (b) The same neutron scattering difference after being symmetrically averaged about the line of symmetry along the (00L) axis of reciprocal space. The data is folded across the (00L) axis and averaged at points where this results in an overlap of the data with itself. (c) The same neutron scattering difference after being symmetrically averaged across the line of symmetry along the (HH0) axis of reciprocal space. (d) The same neutron scattering difference after being fully symmetrized in accordance with both the (00L) and (HH0) lines of symmetry in reciprocal space (also shown in Fig.3(a) of the main letter). This symmetrization brings out relatively weak signals measured in multiple, equivalent {\bf Q} positions and can help minimize non-symmetric features, such as self-absorption effects. (e) A cut through the symmetrized data of (d) along the [H,H,0] direction is shown.  This cut integrates [-0.4,0.4] in the perpendicular [00L] direction.  (f) A cut through the symmetrized data of (d) along the [0,0,L] direction is shown. This cut integrates [-0.4,0.4] in the perpendicular [HH0] direction. }
\end{figure*}

\end{document}